\documentclass[graybox]{svmult}

\usepackage{mathptmx}       
\usepackage{helvet}         
\usepackage{courier}        
\usepackage{type1cm}        
\usepackage{makeidx}         
\usepackage{graphicx}        
\usepackage{multicol}        
\usepackage[bottom]{footmisc}

\usepackage{amssymb}
\usepackage{amsmath}
\usepackage{bm}

\def\beq{\begin{equation}}
\def\eeq{\end{equation}}
\def\nn{\nonumber}
\def\om{\omega}
\def\al{\alpha}
\def\g{\gamma}

\def\s{\sigma}
\def\Del{\Delta}
\def\del{\delta}
\def\ad{a^\dagger}
\def\rd{{\textrm{d}}}
\def\vp{\varphi}
\def\p{\phi}
\def\dd{{\frac{\rd}{\rd z}}}

\def\ra{\rightarrow}

\def\Cc{\mathbb{C}}
\def\Zz{\mathbb{Z}}
\def\Rr{\mathbb{R}}
\def\Zpl{\rm{I\!N}}

\def\Jz{\hat{J}_z}
\def\Jx{\hat{J}_x}
\def\R{\hat{R}}

\def\bz{\bar{z}}

\def\Pr{\hat{P}}
\def\bp{\bar{\phi}}

\def\T{\hat{T}}

\def\Hpm{{\cal H}_\pm}
\def\B{{\cal B}}
\def\hC{\hat{C}}
\def\id{1\!\!\!1}
\def\cI{{\cal I}}
\def\cH{{\cal H}}

\makeindex             

\begin{document}

\title*{Analytical Solutions of Basic Models in Quantum Optics}
\author{Daniel Braak}
\institute{Daniel Braak \at  EP VI and
   Center for Electronic Correlations and Magnetism\\
Institute of Physics, University of Augsburg, Germany\\ \email{daniel.braak@physik.uni-augsburg.de}}
%
%

\maketitle


\abstract{The recent progress in the analytical solution of models invented to  describe theoretically the interaction of matter with light on an atomic scale is reviewed. The methods employ the classical theory of linear differential equations in the complex domain (Fuchsian equations). The linking concept is provided by the Bargmann Hilbert space of analytic functions, which is isomorphic to $L^2(\Rr)$, the standard Hilbert space for a single continuous degree of freedom in quantum mechanics. I give the solution of the quantum Rabi model  in some detail and sketch the solution of its generalization, the asymmetric Dicke model. Characteristic properties of the respective spectra are derived directly from the singularity structure of the corresponding system of differential equations.}


\section{Introduction}

The interaction of matter with light forms a major subject of theoretical and applied physics \cite{cohen-tann}. It is essentially characterized by the quantum nature of both constituents, studied within Quantum Optics \cite{haroche}. The quantum features of the processes occurring in recently realized nano-sized devices can be used to control the generation of entangled states \cite{r-gates}, thereby allowing to construct the basic elements of quantum information technology \cite{nielsen}. 

The technological advances in nanofabrication made it possible to reach very large coupling strengths between the light (usually confined to a single or few modes in a cavity) and an (artificial) atom described by a discrete set of energy levels in the deep quantum limit 
\cite{wallr,niemc,forn}. 
The prototypical system consists of a ``matter'' part with two possible states coupled to the dipole component of a single radiation mode. The Hamiltonian of the atom can thus be expressed via Pauli spin matrices $\s_j$ and the radiation through a harmonic oscillator with frequency $\om$. The Hilbert space of the total system reads then ${\cal H}=\Cc^2\otimes L^2(\Rr)$ and the Hamiltonian 
\beq
H_R= \om\ad a +g\s_x(a+\ad) +\frac{\om_0}{2}\s_z. 
\label{ham1}
\eeq
Here $\ad$ and $a$ are the creation and annihilation operators of the bosonic mode and energy is measured in units of frequency ($\hbar=1$). $\om_0$ denotes the energy splitting of the two-level system (the ``qubit'') which is coupled linearly to the electric field ($\sim (a + \ad)$) with interaction strength $g$. This model was studied semiclassically already in 1936 by Rabi \cite{rabi} and the fully quantized version \eqref{ham1} has been introduced in 1963 by Jaynes and Cummings \cite{JC}. It is therefore called the quantum Rabi model (QRM). Despite its apparent simplicity, the QRM is difficult to solve analytically because it does not exhibit invariant subspaces of finite dimension like the following model,
\beq
H_{JC} = \om\ad a +g(\s^+a + \s^-\ad) +\frac{\om_0}{2}\s_z,
\label{JC}
\eeq
with $\s^\pm = (\s_x\pm i\s_y)/2$, which corresponds to the ``rotating-wave'' approximation of \eqref{ham1} \cite{JC}. The Jaynes-Cummings model (JCM) \eqref{JC} can be justified close to resonance, $\om\sim\om_0$ and small coupling $g/\om\ll 1$ \cite{klimov} 
and has been the standard model for typical quantum optical applications with $g/\om \le 10^{-8}$ for many years. The major simplification arising in \eqref{JC} as compared to \eqref{ham1} consists in the fact that 
the operator $\hC=\ad a +\s^+\s^-$
commutes with $H_{JC}$, which means that $\cal H$ decays into infinitely 
many $H_{JC}$-invariant subspaces in which $\hC$ takes constant values $\in \Zpl_0$, the set of non-negative integers. Each of these spaces is two-dimensional entailing trivial diagonalization of $H_{JC}$.
The fact that the polynomial algebra of $\hC$ has infinite dimension means that $\hC$ generates a continuous $U(1)$-symmetry of $H_{JC}$ \cite{klimov,osterloh}: defining $U(\p)=\exp(i\p\hC)$, we have
$U^\dagger(\p)aU(\p)=e^{i\p}a$, $U^\dagger(\p)\ad U(\p)=e^{-i\p}\ad$ and 
$U^\dagger(\p)\s^\pm U(\p)=e^{\mp i\p}\s^{\pm}$. This abelian symmetry associated with the integral of motion $\hC$ renders the JCM integrable, because it has only two degrees of freedom, the continuous one of the radiation mode with Hilbert space $L^2(\Rr)$ and the discrete one of the qubit with Hilbert space $\Cc^2$. 

The concept of integrability underlying this argument amounts to a direct transfer of Liouville's definition from classical mechanics to quantum mechanics: A system with $N$ degrees of freedom is integrable if it exhibits $N$ independent phase space functions which are in involution with respect to the Poisson bracket \cite{arnold}. In the JCM these are the Hamiltonian $H_{JC}$ and $\hC$. But because independence of operators cannot be defined in analogy to functions on phase space, this definition is not feasible as {\it any} Hamiltonian system would be integrable according to it \cite{caux}. 

The continuous $U(1)$-symmetry of the JCM is broken down to a discrete symmetry
by the counter-rotating term $\ad\s^+ + a\s^-$ in the QRM. 
%
%
$U^\dagger(\pi)H_RU(\pi)=H_R$
 and $H_R$ commutes with $\Pr=(-1)^{\ad a}\s_z=-U(\pi)$.  
Because $\Pr^2=\id$, its polynomial algebra is two-dimensional and $\Pr$ generates a $\Zz_2$-symmetry of $H_R$, usually called parity. The eigenvalues $\pm 1$ of $\Pr$ characterize two $H_R$-invariant subspaces (parity-chains), each of them 
infinite-dimensional \cite{sol1}. Therefore, the problem appears to be only marginally simplified by using the parity symmetry and it was widely held that the QRM is not integrable \cite{osterloh}. However, it could be demonstrated that the weak parity symmetry is indeed sufficient for integrability 
of the QRM because it 
possesses only one {\it continuous} degree of freedom, whereas the Hilbert space dimension of the discrete degree of freedom matches the dimension of the polynomial algebra generated by $\Pr$, rendering the QRM integrable according to the level-labeling criterion for quantum integrability \cite{b-prl}.
The detailed understanding of the spectrum (and dynamics) of the QRM beyond the rotating-wave approximation has been necessitated by the recent experimental access to the ultra-strong and deep-strong coupling regime in circuit QED \cite{niemc,forn} and 
through quantum simulations. 

This survey is organized as follows: In section \ref{sec-QRM}, the analytical solution of the QRM is presented based on a formulation of the problem in Bargmann's space of 
analytical functions, section \ref{sec-dicke} deals with multi-qubit models and the last section contains some remarks on possible future research directions.  
\section{The Quantum Rabi Model}\label{sec-QRM}
The $\Zz_2$-symmetry of the QRM can be used to eliminate the discrete degree of 
freedom from the problem, just as the $U(1)$-symmetry of the JCM allows elimination of 
the continuous degree of freedom. Each parity-chain $\Hpm$ is isomorphic to $L^2(\Rr)$ 
and $H_R$ reads in
$\Hpm$
\beq
H_\pm=\om\ad a +g(a+\ad) \pm\Del(-1)^{\ad a},
\label{hamr}
\eeq
with $\Del=\om_0/2$. The complication of this reduced Hamiltonian comes from the last 
term $(-1)^{\ad a}$. On the other hand, this term is instrumental for the analytical 
solution of the model. To elucidate its meaning, it is convenient to represent \eqref{hamr} in 
Bargmann's space of analytical functions which is isometrically isomorphic to $L^2(\Rr)$ 
\cite{barg}. The space $\B$ is spanned by functions $f(z)$ of a complex variable $z$ 
which have finite norm $\langle \p|\p\rangle$ with respect to the scalar product
\beq
\langle\psi|\phi\rangle =\frac{1}{\pi}\int\rd z\rd \bz e^{-z\bz}\overline{\psi(z)}\phi(z)
\label{b-scalprod}
\eeq
and are analytic in all $\Cc$ ($\rd z\rd\bz=\rd\Re(z)\rd\Im(z)$). The criterion for
 being an element of $\B$ is therefore
two-fold: $\p(z)\in \B$ if both of the following conditions are satisfied:\\
(B-I):\ {$\langle\p|\p\rangle < \infty$}\\
 (B-II):\ {$\p(z)$ is holomorphic everywhere in the open domain $\Cc$.}

The isometry $\cal I$  maps $f(q)\in L^2(\Rr)$ to an analytic function $\p(z)\in \B$,
\beq
\p(z)={\cI}[f](z)=
\frac{1}{\pi^{1/4}}\int_{-\infty}^\infty \!\rd q\ e^{-\frac{1}{2}(q^2+z^2)+\sqrt{2}qz}f(q).
\label{isom}
\eeq
The operators $\ad, a$ are mapped to $z$ and $\rd/\rd z$, respectively,
\beq
\cI a\cI^{-1} = \dd, \qquad \cI\ad\cI^{-1} = z.
\label{isom2}
\eeq  
The normalized vacuum $|0\rangle$ with $a|0\rangle=0$ is mapped 
to the constant function $\p_0(z)=1$. On infers from \eqref{b-scalprod} that all 
polynomials in $z$ are elements of $\B$. 
Especially the $n$-th eigenstate of the harmonic oscillator 
$|n\rangle\sim e^{-q^2/2}H_n(q)$ is mapped onto the monomial $z^n/\sqrt{n!}$.
Moreover, all functions which have the 
asymptotic expansion
\beq
\p(z)=e^{\al_1 z}z^{-\al_0}(c_0+c_1z^{-1}+c_2z^{-2}+\ldots) \quad \textrm{for} \quad z\ra \infty,
\label{asym-1}
\eeq
with arbitrary $\al_1\in\Cc$ satisfy (B-I) \cite{b-contfrac}.
\eqref{asym-1} is the asymptotic form of the normal solutions of a differential equation having an unramified irregular singular point of s-rank two at infinity \cite{ince,slav}. 
The coherent state $|\al\rangle=e^{\al\ad}|0\rangle$ in $L^2(\Rr)$ is mapped obviously 
to the exponential function $e^{\al z}$.
Functions behaving as
\beq
\p(z)=\exp\left(\frac{\al_2}{2}z^2+\al_1 z\right) z^{-\al_0}(c_0+c_1z^{-1}+c_2z^{-2}+\ldots) 
\label{asym-2}
\eeq
asymptotically, satisfy (B-I) only if $|\al_2|<1$ and correspond to an irregular singularity of s-rank three. The limiting value
$|\al_2|=1$  belongs to the (not normalizable) plane wave states 
$f_p(q)=\exp(ipq)/\sqrt{2\pi}$,
\beq
\cI[f_p](z)= \frac{e^{-p^2/2}}{\pi^{1/4}}\exp\left(\frac{1}{2}z^2 +i\sqrt{2}pz\right).
\label{planewaves}
\eeq
The Hamiltonian $H_+$ reads in $\B$ (with $\om=1$),
\beq
H_+=z\dd +g\left(z+\dd\right) +\Del\T,
\label{hamb}
\eeq
where $\T$ denotes the reflection operator $\T[\psi](z)=\psi(-z)$.
The Schr\"odinger equation $(H_+-E)\psi(z)=0$ corresponds to a linear but non-local 
differential equation in the complex domain,
\beq
z\dd\psi(z) +g\left(\dd + z\right)\psi(z) = E\psi(z) -\Del\psi(-z).
\label{time-i-schr}
\eeq
The theory of these equations initiated by Riemann and Fuchs \cite{ince} can now be 
applied to \eqref{time-i-schr}. First, one obtains with the 
definition $\psi(z)=\p_1(z)$ and $\psi(-z)=\p_2(z)$ the coupled local system,
\begin{subequations}
\label{coup-sys}
\begin{eqnarray}
(z+g)\dd\p_1(z) +(gz-E)\p_1(z) +\Del\p_2(z) &=& 0,\\
(z-g)\dd\p_2(z) -(gz+E)\p_2(z) +\Del\p_1(z) &=& 0. 
\end{eqnarray}
\end{subequations}
This system has two regular singular points at $z=\pm g$ and an (unramified) 
irregular singular point
of s-rank two at $z=\infty$ \cite{slav}. The normal solutions of \eqref{coup-sys} 
behave asymptotically as \eqref{asym-1} with $\al_1=g$ or $\al_1=-g$. The two Stokes rays 
are the positive and negative real axis. We infer that {\it all} solutions 
of \eqref{coup-sys} satisfy (B-I). I follows that the discrete 
spectrum $\{E_n\}$, $n\in\Zpl_0$ will be determined by (B-II), 
because not all solutions of \eqref{coup-sys} are analytic in $\Cc$. 
Define $x=E+g^2$. Then
the exponents of $\p_1(z)$ at the regular singular point $g$ ($-g$) 
are $\{0,1+x\}$ ($\{0,x\}$), while for $\p_2(z)$ 
the exponents at  $g$ ($-g$) are $\{0,x\}$ ($\{0,1+x\}$).  

If $E$ belongs to the spectrum of $H_+$, $\p_1$ and $\p_2$ must be analytic 
in $\Cc$, especially at both points $\pm g$. This leads naturally to a 
division of the spectrum into two parts:\\
1)\ The {\it regular} spectrum $\s_{reg}$ consisting of those values $E_n$ for which 
$x_n=E_n+g^2$ is not a non-negative integer.\\
2)\ The {\it exceptional} spectrum $\s_{exc}$ for which $x_n\in\Zpl_0$.
\subsection{The Regular Spectrum}\label{subsec-reg}
\begin{figure}[h]
\sidecaption
    \includegraphics[scale=.48]{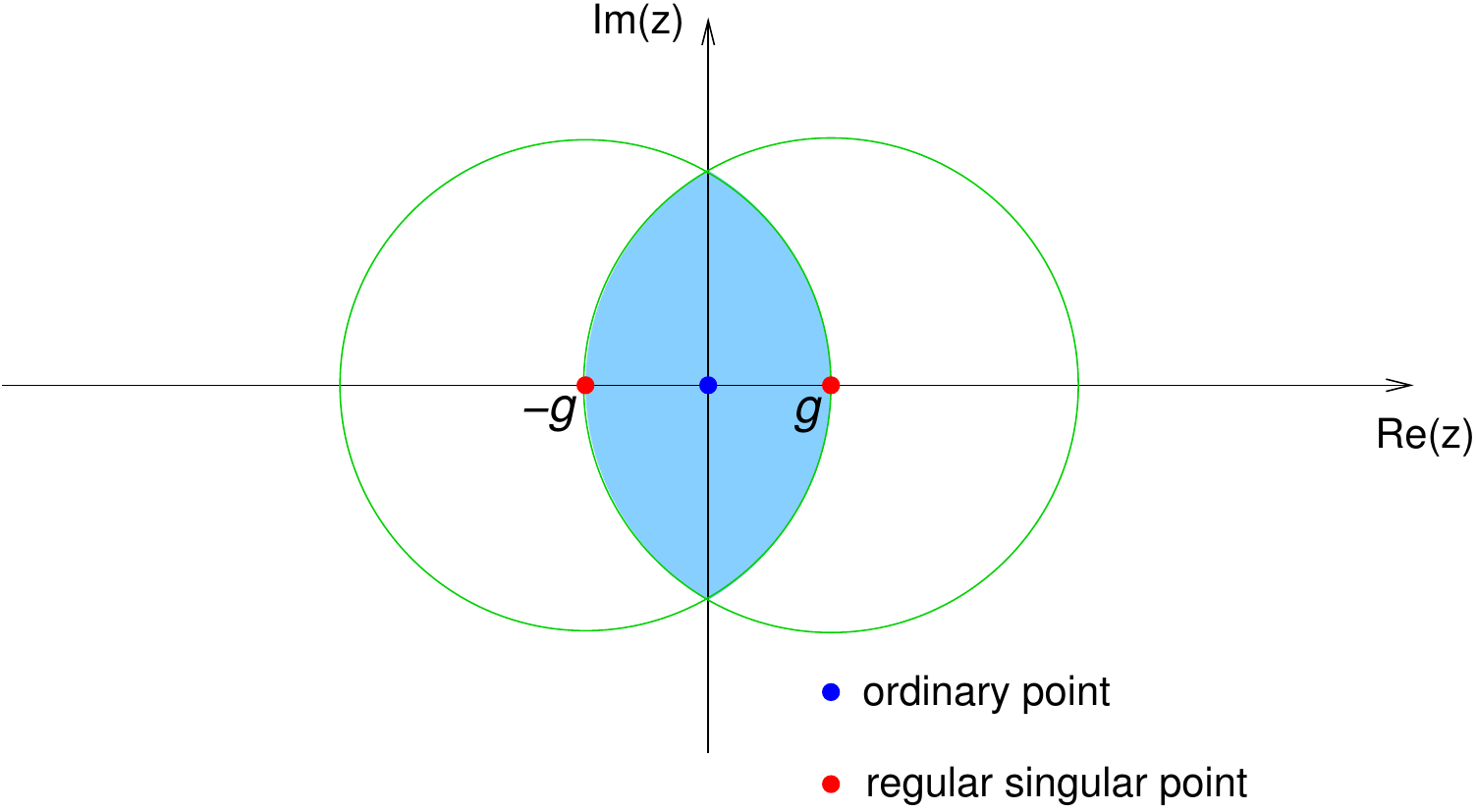}
   \caption{The singularity structure of \eqref{time-i-schr} and \eqref{coup-sys}. 
Two local Frobenius solutions analytic at $z=g$ and $z=-g$ respectively are 
defined by \eqref{recur}, \eqref{f-n}. 
If they coincide in the intersection of their  domains of convergence, 
they describe the same (analytic) function in $\Cc$.
   }
    \label{sing-rabi}
  \end{figure}

\begin{figure}[h]
    \sidecaption
    \includegraphics[scale=.6]{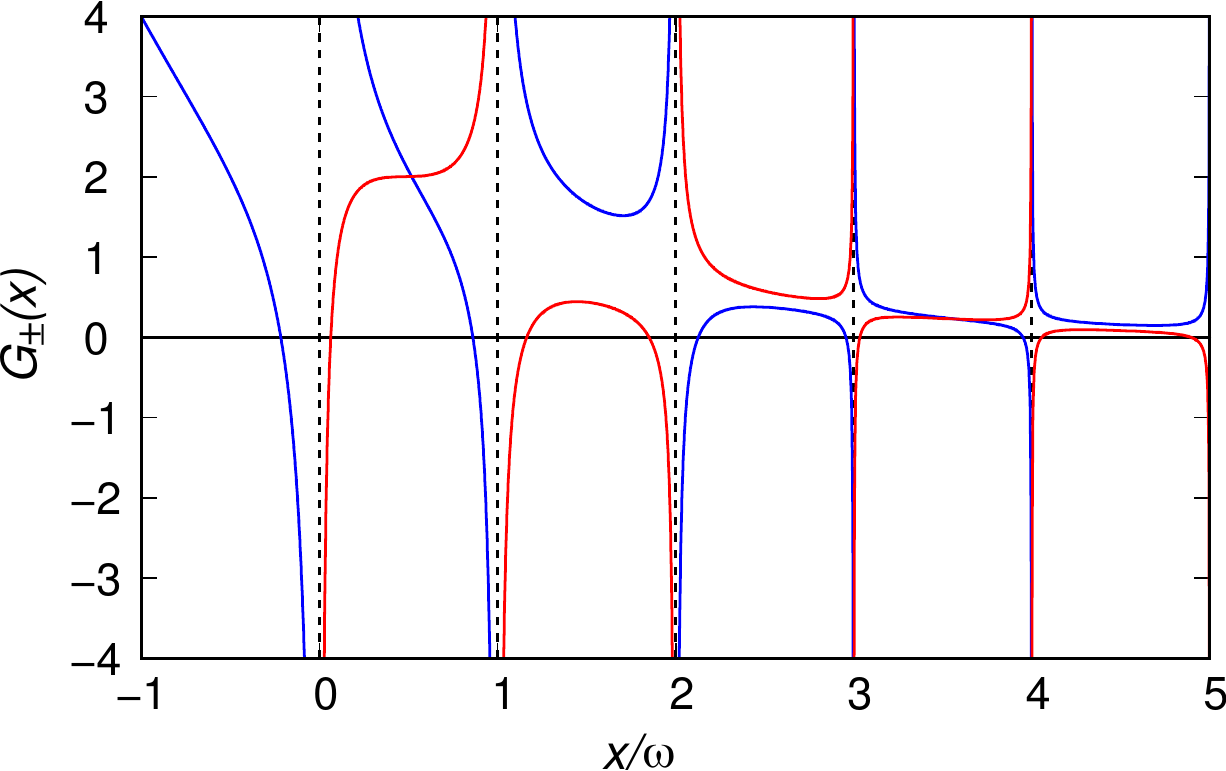}
\caption{The $G$-functions for odd (blue) and even (red) 
parity for $g=\om=1$ and $\Del=0.4$.}
\label{Gfunction}
\end{figure}

\begin{figure}[b]
\centering
    \includegraphics[scale=.6]{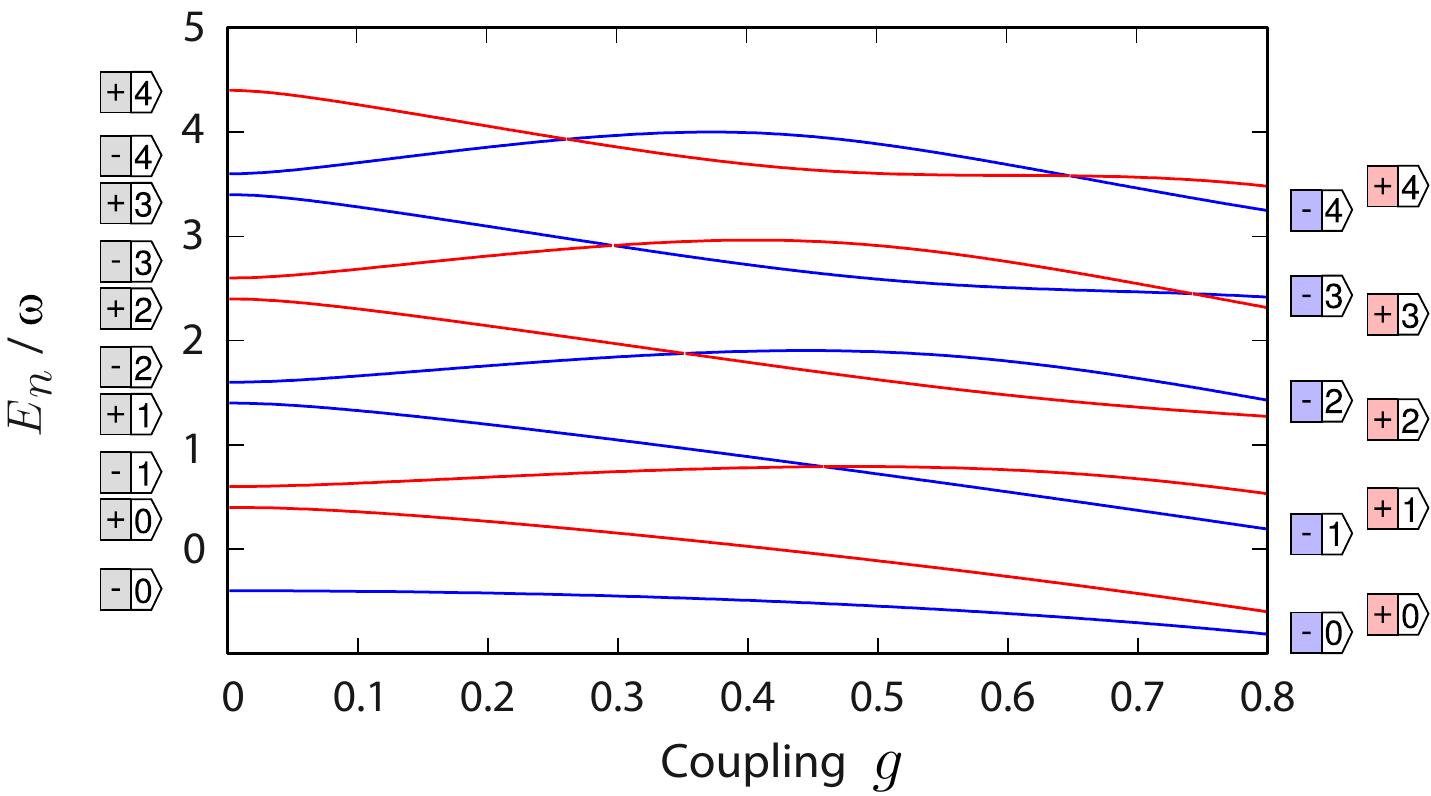}
   \caption{ Rabi spectrum for the same 
$\Del,\om$ as in Fig.~\ref{Gfunction} and  $0\le g \le 0.8$. The intersections between the spectra of 
different parity indicate the degenerate part of the exceptional spectrum. 
The two-fold labeling of states on the left corresponds to the uncoupled 
system (the $\pm$ denotes the spin quantum number) and on the right to 
the coupled case ($\pm$ denotes the parity quantum number).}
    \label{spectrum}
  \end{figure}

If $x\notin\Zpl_0$, the only allowed exponent at both points $\pm g$ is 0. 
We consider the case $-g$ and define $y=z+g$, $\p_{1,2}=e^{-gy+g^2}\bp_{1,2}$. Then,
\begin{eqnarray}
\phantom{(y-2)}y\frac{\rd}{\rd y}\bp_1 &=& x\bp_1-\Del\bp_2,
\label{coup-sys2}\\
(y-2g)\frac{\rd}{\rd y}\bp_2 &=& (x-4g^2+2gy)\bp_2 -\Del\bp_1.
\label{coup-sys3}
\end{eqnarray}
A local Frobenius solution for $\bp_2(y)$, analytic at $y=0$, 
reads $\bp_2(y)=\sum_{n=0}^\infty K_n(x)y^n$ with coefficients $K_n(x)$ to be determined. 
Integration of \eqref{coup-sys2} yields
\beq
\bp_1(y)=cy^x-\Del\sum_{n=0}^\infty K_n(x)\frac{y^n}{n-x}.
\label{bp1}
\eeq  
Because $x\notin\Zpl_0$, $c$ must be zero. This determines $\bp_1(z)$ uniquely in terms of
$\bp_2(z)$. Setting $K_0=1$, the following three-term recurrence relation for the $K_n(x)$
is obtained  from \eqref{coup-sys3},
\beq
nK_{n}=f_{n-1}(x)K_{n-1}-K_{n-2},
\label{recur}
\eeq
with
\beq
f_n(x)=2g+\frac{1}{2g}
\left(n-x +\frac{\Del^2}{x-n}\right),
\label{f-n}
\eeq
and initial condition $K_0=1, K_1(x)=f_0(x)$. $\bp_2(y)$ will usually develop 
a branch-cut singularity at $y=2g$. The radius of convergence of the Frobenius 
solution around $y=0$ is $R=2g$, which can be deduced from the asymptotic 
value $1/(2g)$ of $f_{n-1}(x)/n$ for 
$n\ra\infty$. Due to the relation $\p_2(z)=\psi(-z)$, the formal solution of 
\eqref{coup-sys2} and \eqref{coup-sys3} yields two expansions for $\psi(z)$, 
one analytic at $z=g$ and the other analytic at $z=-g$,
\begin{eqnarray}
\psi(z)=\p_2(-z)&=& e^{gz}\sum_{n=0}^\infty K_n(x)(-z+g)^n \label{psi-2},\\
\psi(z)=\p_1(z) &=& e^{-gz}\sum_{n=0}^\infty K_n(x)\Del\frac{(z+g)^n}{x-n}.
\label{psi-1}
\end{eqnarray}
The two circles centered at $z=\pm g$ are shown in Fig.~\ref{sing-rabi}. 
%
%
Because the vectors $(\p_1(z),\p_2(z))^T$ and $(\p_2(-z),\p_1(-z))^T$ 
satisfy both the homogeneous first-order system \eqref{coup-sys}, they 
coincide in a neighborhood of $z_0$ if 
\beq
\p_1(z_0)=\p_2(-z_0), \quad \p_1(-z_0)=\p_2(z_0)
\label{condition}
\eeq 
for any $z_0$ in the intersection of their domain of convergence. That means that 
$\p_2(-z)$ is the analytic continuation of $\p_1(z)$ and itself 
analytic at $z=g$, therefore 
$\psi(z)$ is analytic at both singular points.
Both conditions in
\eqref{condition} are equivalent if $z_0=0$ \cite{b-wronski}. This leads 
to the definition of the
$G$-function for the regular spectrum of $H_+$ \cite{b-prl,zhong},
\beq
G_+(x)=\p_2(0)-\p_1(0)=\sum_{n=0}^\infty K_n(x)
\left[1-\frac{\Del}{x-n}\right]g^n.
\label{G-reg}
\eeq
If $G_+(E_n+g^2)=0$, the corresponding formal solution $\psi(z)$ is analytic 
everywhere and an element of $\B$ because it satisfies (B-I) and (B-II), entailing 
that $E_n\in \s_{reg}(H_+)$.  $G_-(x)$ for $H_-$ is obtained from $G_+(x)$ by 
replacing $\Del$ with $-\Del$ in \eqref{G-reg}. It follows 
from \eqref{recur}, \eqref{f-n} and \eqref{G-reg} that $G_\pm(x)$ has simple poles at
$x\in\Zpl_0$. The zeros of $G_\pm(x)$ are distributed between these poles. 
Fig.~\ref{Gfunction} shows $G_\pm(x)$ and Fig.~\ref{spectrum} the corresponding spectrum of $H_R$ for both parities.
%
%
It is easy to see that the regular spectrum is never degenerate, neither 
within each parity chain nor among states with different parity \cite{b-prl}.
The $G$-function can be expressed in terms of known special functions as follows
\beq
G_\pm(x)=
\left(1\mp\frac{\Del}{x}\right)H_c(\al,\g,\del,p,\s;1/2)
-\frac{1}{2x}H'_c(\al,\g,\del,p,\s;1/2).
\label{G}
\eeq
$H_c(\al,\g,\del,p,\s;z)$ denotes a confluent Heun-function \cite{ronv} and 
$H'_c(\al,\g,\del,p,\s;z)$ its derivative with respect to $z$. 
The parameters are given as 
\begin{align}
\al&=-x, \quad \g=1-x, \quad \del=-x,\nn\\
p&=-g^2,\quad \s=x(4g^2-x)+\Del^2. \nn
\end{align}
The functional form \eqref{G-reg} of $G_\pm(x)$ leads to the following 
conjecture about the distribution of its zeros along the positive real axis.\\
{\bf Conjecture}:\ The number of zeros in each interval $[n,n+1]$, $n\in\Zpl_0$
is restricted to be 0, 1, or 2. Moreover,  an
interval $[n,n+1]$ with two roots of $G_\pm(x)=0$ can 
only be adjacent to an interval
with one or zero roots; in the same way, an empty interval can never
be adjacent to another empty interval. 
\subsection{The Exceptional Spectrum}\label{subsec-exc}
We shall demonstrate in the following that the presence of the 
exceptional spectrum $\s_{exc}$ poses certain constraints on the 
model parameters $g$ and $\Del$ such that for given $g$, $\Del$, at 
most two eigenvalues are exceptional. 
Furthermore, $\s_{exc}=\s_{exc}^d \cup \s_{exc}^{nd}$, where $\s_{exc}^d$ comprises
the values $E=m-g^2$ with $m\in\Zpl$. Each eigenvalue in $\s_{exc}^d$ 
is doubly degenerate among states with different parity. $\s_{exc}^{nd}$ is not degenerate
and may take values $E=m-g^2$ with $m\in\Zpl_0$.
 
We begin with $\s_{exc}^{nd}$. The poles of $G_+(x)$ at $x\in \Zpl_0$ 
indicate that an integer $x$ can only signify an eigenvalue of $H_+$ 
if the corresponding pole in $G_+(x)$ is lifted for special values 
of the parameters $g$ and $\Del$. If $x\in \Zpl_0$, not only the 
exponent 0 but also the exponents $x$, respectively $x+1$ guarantee 
analyticity of $\p_1(z)$ and $\p_2(z)$ at $z=-g$. However, as the 
difference of the two exponents at both singular points is a 
positive integer if $x>0$ (for $x=0$ this difference is positive 
at one singular point), the local 
analytic Frobenius solutions around $z=-g$ will develop a 
logarithmic branch-cut at $z=g$ in general. For $x=m\in\Zpl_0$, there exist 
always a solution for $\bp_2(y)$ analytic at $y=0$ of the form
\beq
\bp_2(y)=\sum_{n=m+1}^\infty K_ny^n,
\label{bp2-ex}
\eeq
because the largest exponent of $\bp_2(y)$ at $y=0$ is $x+1$ \cite{ince}. 
Integration of \eqref{coup-sys2} yields for $\bp_1(y)$,
\beq
\bp_1(y)=cy^m-\Del\sum_{n=m+1}^\infty K_n\frac{y^n}{n-m}.
\label{bp1-ex}
\eeq
In this case, the constant $c$ may be different from zero because $\bp_1(y)$ 
is then analytic at $y=0$. Solving now \eqref{coup-sys3} with the 
ansatz \eqref{bp2-ex}, we obtain for 
$n\ge m+2$ the recurrence \eqref{recur}, \eqref{f-n} and the initial conditions
\beq
K_{m+1}=\frac{c\Del}{2(m+1)g},\qquad K_m=0.
\eeq
$c$ is fixed in terms of $K_{m+1}$. Setting $K_{m+1}=1$ we obtain 
for $\psi(z)$ the two expressions
\begin{eqnarray}
\psi(z)&=&\p_2(-z)= e^{gz}\sum_{n=m+1}^\infty K_n(m;g,\Del)(-z+g)^n, \label{psi-2ex}\\
\psi(z)&=&\p_1(z) = e^{-gz}\left(\frac{2(m+1)g}{\Del}(z+g)^m
-\Del\hspace{-2mm}\sum_{n=m+1}^\infty \hspace{-1mm}K_n(m;g,\Del)\frac{(z+g)^n}{n-m}\right)
\label{psi-1ex}
\end{eqnarray}
and the $G$-function follows as  
\beq
 G^{(m)}_+(g,\Del)=-\frac{2(m+1)}{\Del}+\sum_{n=m+1}^\infty 
K_n(m;g,\Del)\left(1+\frac{\Del}{n-m}\right)g^{n-m-1}.
\label{G-exc}
\eeq
The zeros of the function $G^{(m)}_+(g,\Del)$ determine those values of the parameters
$g$ and $\Del$ for which $H_+$ has the exceptional eigenvalue $m-g^2$ with $m\in\Zpl_0$.
For odd parity, we have  $G^{(m)}_-(g,\Del)=G^{(m)}_+(g,-\Del)$. It follows 
that $G^{(m)}_+(g,\Del)$ and   $G^{(m)}_-(g,\Del)$ have no common zeros, so 
this part of the exceptional spectrum is not degenerate, just as the 
regular spectrum. It was computed by a related method in
\cite{mac}. 

To obtain $\s_{exc}^d$, we consider now the smaller exponent, zero, 
of $\bp_2(y)$ at $y=0$, leading to the expansion
 \beq
\bp_2(y)=\sum_{n=0}^\infty K_ny^n.
\label{bp2-judd}
\eeq
After integration of \eqref{coup-sys2}, $\bp_1(y)$ reads
\beq
\bp_1(y)=cy^m-\Del\sum_{n\neq m}^\infty K_n\frac{y^n}{n-m} -\Del y^mK_m\ln(y).
\label{bp1-judd}
\eeq
The $K_n$ for $n\le m-1$ are determined again with \eqref{recur} and initial conditions
$K_0=1$, $K_1=f_0(m)$. Therefore $K_m(m;g,\Del)$ is uniquely fixed. 
The logarithmic term in \eqref{bp1-judd} vanishes if  $K_m(m;g,\Del)=0$ 
\cite{b-prl}.
The coefficients $K_n$ for $n\ge m+1$ are determined with \eqref{recur}
and initial conditions
\[
K_{m+1}=\frac{1}{m+1}\left[\frac{c\Del}{2g}-K_{m-1}\right], \qquad K_m=0.
\]
In this case there exist two local solutions analytic at $y=0$, 
\eqref{bp2-ex} and \eqref{bp2-judd}. If $m\neq 0$, they are linearly 
independent and span the whole solution space for $\bp_2(y)$. 
Because of the reflection symmetry mapping $y$ to $2g-y$, 
these solutions describe the solution space in 
a neighborhood of $y=2g$ as well and {\it all} solutions of \eqref{coup-sys}
are analytic at both $g$ and $-g$, thus in all of $\Cc$, if  $K_m(m;g,\Del)=0$ 
and no further condition is necessary. Moreover,  
the $K_m$'s are the same for odd parity, so the eigenvalue $E=m-g^2$ obtained 
via this condition is always doubly degenerate between states of different 
parity. 
The presence of the spectrum $\s_{exc}^d$ and its ``quasi-exact'' nature \cite{turb} may be explained more generally in terms of the representation theory of $\mathfrak{sl}_2(\Rr)$ \cite{wak1}.
A special situation arises for $x=0$. The condition $K_0(0)=0$ 
renders the ansatz \eqref{bp2-judd} equivalent to
\eqref{bp2-ex} and both solutions are linearly dependent. Thus only one 
local solution may be analytic at $y=0$ and
$\bp_1(y)$ is given by \eqref{bp1-ex} for $m=0$. If $E=-g^2$ is an 
eigenvalue of $H_+$, $G^{(0)}_+(g,\Del)$ must vanish. It follows that for these parameter values
 $G^{(0)}_-(g,\Del)\neq 0$, the eigenvalue $E=-g^2$ is never degenerate 
and an element of $\s_{exc}^{nd}$.
\subsection{Methods based on Continued Fractions}\label{subsec-contfrac}
The Bargmann space formalism  has been applied to the QRM as early as 
1967 by Schweber \cite{schweb}. He did not make use of the $\Zz_2$-symmetry 
but derived the coupled system \eqref{coup-sys} directly from \eqref{ham1} with the ansatz
$\bm{\psi}=(\vp_1(z),\vp_2(z))^T\in \B\otimes\Cc^2$ for the wave function 
with energy $E$. \eqref{coup-sys} is then satisfied by $\p_1=\vp_1+\vp_2$ and
$\p_2=\vp_1-\vp_2$. He obtained the local Frobenius solution for $\bp_2(y)$ 
given by \eqref{recur} and \eqref{f-n}. The convergence radius of the series
\eqref{bp2-judd} is $2g$ for arbitrary $x$ and the discrete set of 
eigenvalues is selected by determining those $x$ for which 
\eqref{bp2-judd} has infinite convergence radius. The problem is 
equivalent to compute the {\it minimal solution}
of the recurrence \eqref{recur} \cite{gautschi}. The spectral 
condition obtains then by equating the minimal $K_1(x)/K_0$  
with $f_0(x)$ from the initial conditions. The equation has 
the form $F(x)=0$, where $F(x)$ is represented by a continued fraction \cite{schweb}.
This method, while formally correct, has several conceptual shortcomings:
\begin{itemize}
\item The function $F(x)$ has an unknown singularity structure and it is impossible to 
infer qualitative aspects on the distribution 
of its zeros from it.
\item The actual computation of the continued fraction makes a truncation at some order 
necessary which is equivalent to  define the model on a finite-dimensional Hilbert 
space, which is the starting point for other work employing continued fractions 
to compute the Rabi spectrum \cite{swain,tur}.
\item The zeros of $F(x)$ correspond to  $\s_{reg}\cup\s_{exc}^{d}$, but there is 
no possibility to discern both types of spectra, especially the double 
degeneracy of $\s_{exc}^d$ cannot be detected with this method.
\item The spectrum $\s_{exc}^{nd}$ is not accessible because the expansion
\eqref{bp2-judd} with $K_0\neq 0$ is assumed in the derivation of $F(x)$.
\end{itemize} 

Besides these conceptual problems, the method is numerically feasible 
only for the first low-lying eigenvalues. The continued fraction has a 
pole in close vicinity of each zero and their distance approaches zero exponentially  
for growing $x$, so that at most ten energy levels may be resolved 
within a double precision calculation. On the other hand, 
the equivalence of the continued fraction approach to exact 
diagonalization in finite-dimensional Hilbert spaces proves 
the validity of the latter for the QRM \cite{b-contfrac}. 

Schweber's technique is confined to problems reducible to three-term 
recurrence relations for the local Frobenius solutions and 
implements then (B-II) as the spectral condition. It fails for 
models with more than a single qubit because the ensuing recurrence relations have more than three terms. 
The next section is devoted to the application of the theory 
presented above to models with $N>1$ qubits. 
\section{The Dicke Models}\label{sec-dicke}
The natural generalization of the Hamiltonian \eqref{ham1} couples several two-level systems to the same mode of the radiation field,
\beq
H_{DN}=\om\ad a +\sum_{i=1}^N\frac{\om_{0i}}{2} \s_{iz} + (a+\ad)\frac{1}{\sqrt{N}}\sum_{i=1}^N g_i'\s_{ix}.
\label{hamd1}
\eeq
This model assumes different qubit frequencies $\om_{0i}$ and couplings $g_i$
to the field and is therefore called the asymmetric Dicke model (ADM$_N$) with $N$ qubits. Dicke introduced the permutation-invariant version of \eqref{hamd1} in 1954 and predicted the (later observed) phenomenon of ``superradiance'' for large $N$ \cite{dicke}. Its rotating-wave approximation is integrable for all $N$ \cite{tavis}, while the full model is non-integrable for any $N>1$ according to the level-labeling criterion \cite{b-prl}.  
Applications in quantum information technology mandate the study of 
\eqref{hamd1} without approximations for small $N$, because it describes  the implementation of quantum gates within circuit QED \cite{haack}.

The following section treats the asymmetric model for $N=2$ and section
\ref{sec-d3} the symmetric model for $N=3$.
\subsection{ADM$_2$ and Exceptional States}\label{sec-d2}
The Hamiltonian of the ADM$_2$ reads in slightly different notation,
\beq\label{hamd2}
H_{D2}=\omega
a^{\dagger}a+g_{1}\sigma_{1x}(a+a^{\dagger})+g_{2}\sigma_{2x}(a+a^{\dagger})
+\Delta_1\sigma_{1z}+\Delta_2\sigma_{2z}.
\eeq
This model has a $\Zz_2$-symmetry similar to \eqref{ham1}, generated
by $\Pr=\exp(i\pi\ad a)\s_{1z}\s_{2z}$. However, because it has only two irreducible one-dimensional representations, the discrete degrees of freedom cannot be labeled with a $\Zz_2$-quantum number (their Hilbert space is $\Cc^4$). The symmetry of the model is not sufficient to make it integrable as the $N=1$ case \eqref{ham1}.

Nevertheless, the same methods as above can be used to solve \eqref{hamd2} exactly \cite{peng}, at the expense of a more complicated $G$-function, which is no longer a linear combination of formal solutions as \eqref{G-reg}. After application of the symmetry, the remaining Hilbert space is not $\B$, but $\B\otimes\Cc^2$. The Hamiltonian reads ($\om=1$),
\beq
H_\pm=z\dd + (g_1+g_2\s_z)\left(z+\dd\right) + (\Del_2\pm\Del_1\T)\s_x,
\label{hamrd2}
\eeq
with $\T=(-1)^{z\dd}$.
An eigenfunction of \eqref{hamrd2} with eigenvalue $E$ is the vector
$\bm{\psi}=(\varphi_1(z),\varphi_2(z))^T$. Defining  $\varphi_{3}(z)=\varphi_{1}(-z)$ and $\varphi_{4}(z)=\varphi_{2}(-z)$, we obtain a coupled system of four ordinary first order differential equations with four regular singular points located at $g=g_1+g_2$, $g'=g_1-g_2$, $-g$ and $-g'$.
Moreover, $z=\infty$ is an unramified irregular singular point of s-rank two; we conclude that again all formal solutions fulfill (B-I) and $E$ is determined by postulating analyticity of the solution at all regular singular points.  The corresponding coupled recurrence relations for the Frobenius expansions around each of the points $0$, $g$ and $g'$ cannot be reduced to a three-term recurrence, except in the case $g'=0$, which allows a treatment similar to the QRM. It turns out that eight initial conditions determine functions $\phi_k(z)$, $k=1,\ldots,32$, describing the $\varphi_j(z)$ around different expansion points. Overall analyticity is then equivalent to the vanishing of the determinant of a $8\times 8$-matrix $M_\pm(E)$, whose entries are composed of the $\phi_k$, evaluated at the ordinary points $z_0$ and $z_0'$, whose location depends on the geometry of the analytic regions \cite{peng}. The $G$-function can then be defined as $G_\pm(E)=\det(M_\pm(E))$.
This function has poles at integer values of $E+{g'}^2$ and $E+g^2$, defining the exceptional spectrum, besides the regular, given by the condition
$G(E_n)=0$.

\begin{figure}[h]
    \centering
    \includegraphics[scale=0.35]{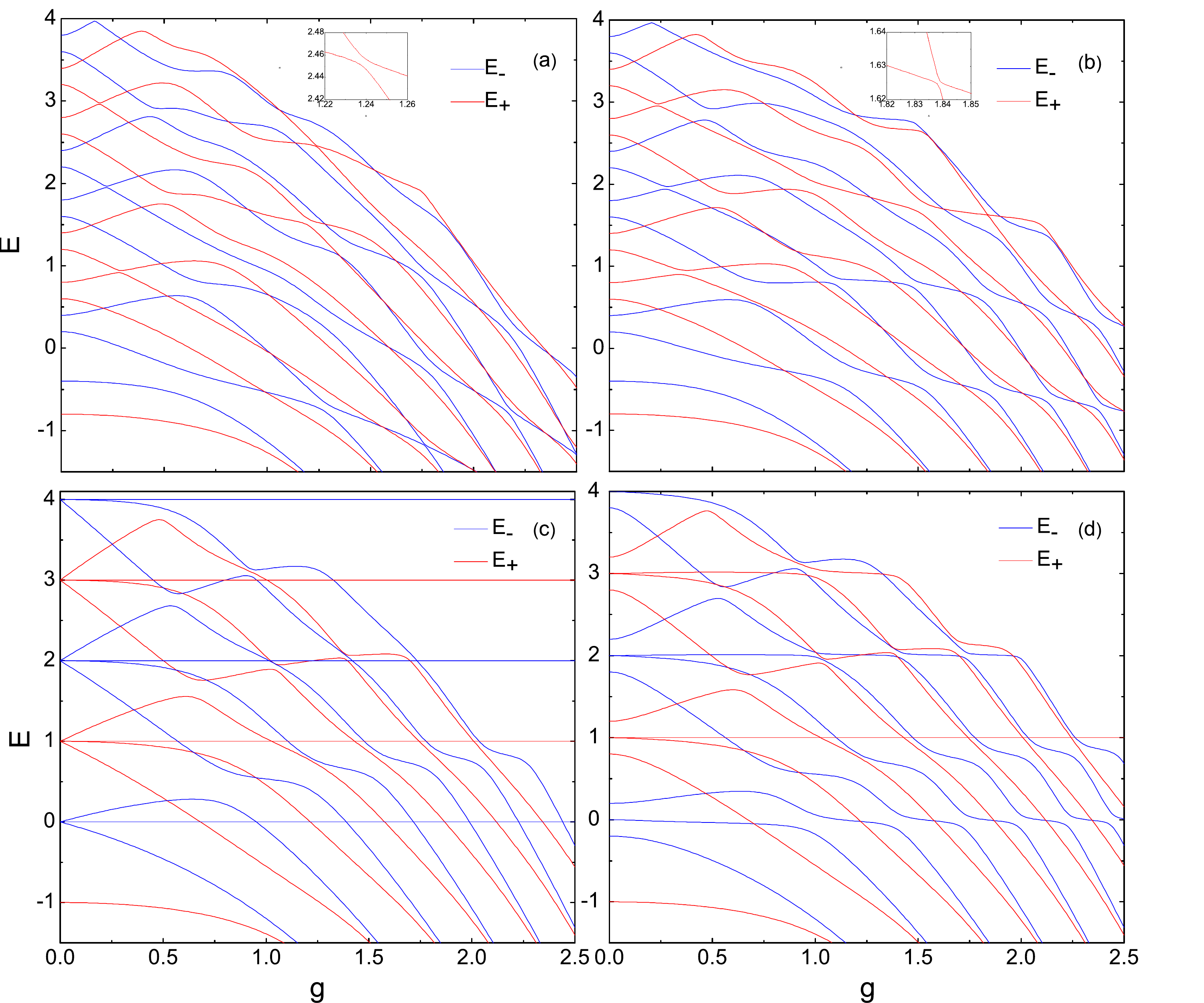}
   \caption{The spectra of $H_{D2}$ with
(a) $\Delta_1=0.6$,~$\Delta_2=0.2$,~$\omega=1$,~$0\leq
g=g_1+g_2\leq2.5$,~$g_1=4g_2$.~ (b)
$\Delta_1=0.6$,~$\Delta_2=0.2$,~$\omega=1$,~$0\leq
g=g_1+g_2\leq2.5$,~$g_1=2g_2$.~(c)$\Delta_1=\Delta_2=0.5$,~$\omega=1$,~$0<g=g_1+g_2<2.5$,~$g_1=g_2$.
 ~(d)$\Delta_1=0.6$,~$\Delta_2=0.4$,~$\omega=1$,~$0\leq g\leq
2.5$,~$g_1=g_2$. Blue lines are eigenvalues with odd
parity, while red  lines are eigenvalues with even
parity.}
    \label{dicke2}
  \end{figure}

The spectra of $H_{D2}$ obtained in this way are depicted in Fig.~\ref{dicke2}, as function of $g$ and various levels of asymmetry.
%
%
For the completely asymmetric cases in Figs.~\ref{dicke2}(a), \ref{dicke2}(b), we observe level crossings between states of different parity, whereas states with equal parity show avoided crossings, some of them quite narrow as the insets demonstrate. There is no relation between degeneracies and the exceptional spectrum as in the case of the QRM, because the $G$-functions $G_+(E)$ and $G_-(E)$ are not simply related. The set of lines with $E+g^2=n$, $n\in\Zpl_0$, give the asymptotic (parity degenerate) spectrum in the deep strong coupling limit.  Fig.\ref{dicke2}(c) shows the
completely symmetric case, $\Del_1=\Del_2$, $g_1=g_2$. The invariance under permutation symmetry of \eqref{hamd2} leads to separation of the Hilbert space of the spin-1/2 qubits into singlet and triplet sector according to $\frac{1}{2}\otimes\frac{1}{2}=0\oplus 1$. The total Hilbert space becomes thus
$\B\otimes\Cc^4\rightarrow\B\oplus\B\otimes\Cc^3$. The singlet subspace is isomorphic to $\B$ and the Hamiltonian describes the decoupled radiation mode. Thus the eigenenergies are just integer multiples of $\om$, independent of the coupling. They are seen as horizontal lines in Fig.~\ref{dicke2}(c). The triplet subspace is coupled to the radiation field and the spectrum shows a nontrivial dependence on $g$. 

An interesting situation obtains for equal couplings $g_1=g_2$ but different qubit energies, $\Del_1\neq\Del_2$. The full permutation symmetry is broken, but there is a certain remnant of it. For $g_1=g_2$, $z=0$ is a regular singular point and there exist ``quasi-exact'' eigenstates, belonging to the exceptional spectrum with $E_n\in \Zpl$ for certain parameter values of $g$, $\Del_j$. These states contain a finite number $N$ of photons (contrary to the likewise quasi-exact elements of $\s_{exc}^{d}$ in the QRM) and are determined by a polynomial equation for $g,\Del_1,\Del_2$ depending on the energy value $E=N$. It reads for $N=2$,
\beq
\left[\left(2-\frac{(\Del_2\pm\Del_1)^2}{2}\right)\big(1-(\Del_2\mp\Del_1)^2\big)-g^2\right](\mp\Del_1-\Del_2)=0,
\eeq
where the $+$ ($-$) corresponds to even (odd) parity. The condition comprises the symmetric case $\Del_1=\Del_2$ for odd parity, but otherwise determines $\Del_1$, $\Del_2$
in terms of $g$. This is true for all $N\ge 2$. However, for $N=1$ we find,
\beq
(\pm\Del_1-\Del_2)\big[1-(\Del_2\pm\Del_1)^2\big]=0.
\label{Neq1}  
\eeq
There exists an eigenstate with energy $E=1$ and even (odd) parity, if $\Del_1+\Del_2=1$, ($|\Del_1-\Del_2|=1$). This state contains at most one photon and the condition for its existence does {\it not} depend on $g$. 
It was first discovered by Chilingaryan and Rodr\'iguez-Lara \cite{rod}. With the notation
$|\psi\rangle=|n,s_1,s_2\rangle$ for a basis element in $\cH$, where we have used the occupation number basis for the boson mode and $s_j\in\{g,e\}$ denotes the state of the $j$-th qubit, the exceptional state with even parity ($\Del_1+\Del_2=1$) reads,
\beq
|\psi_e\rangle=\frac{1}{\cal N}\left(\frac{2(\Del_1-\Del_2)}{g}|0,e,e\rangle-|1,e,g\rangle+|1,g,e\rangle\right),
\label{Neq1-state}
\eeq
with a normalization factor $\cal N$. This state becomes the singlet state
\beq
\frac{1}{\sqrt{2}}|1\rangle\otimes(|g,e\rangle-|e,g\rangle)
\label{singlet}
\eeq
in the symmetric case $\Del_1=\Del_2$. The spectrum for $\Del_1+\Del_2=1$ is shown in 
Fig.\ref{dicke2}(d), exhibiting the quasi-exact state with $E=1$ as a $g$-independent line. In contrast to the fully decoupled singlet state \eqref{singlet}, the state \eqref{Neq1-state} is strongly coupled to the radiation field, as its components depend on $g$. It is quite remarkable that states with finite maximal photon number exist for arbitrary strong coupling without making the rotating wave approximation. This feature cannot be realized in the QRM, where each eigenstate contains always an infinite number of photons. Due to its very simple structure, the state \eqref{Neq1-state} could be useful for quantum computing applications, especially as the condition for its existence depends only on the (easily controllable) qubit energies $\Del_1$ and $\Del_2$ and not on the coupling strength. Similar states are expected to exist in all models ADM$_N$ with $g_i\equiv g$ and even $N$. 
\subsection{ADM$_3$}\label{sec-d3}
\begin{figure}[h]
    \centering
    \includegraphics[scale=0.6]{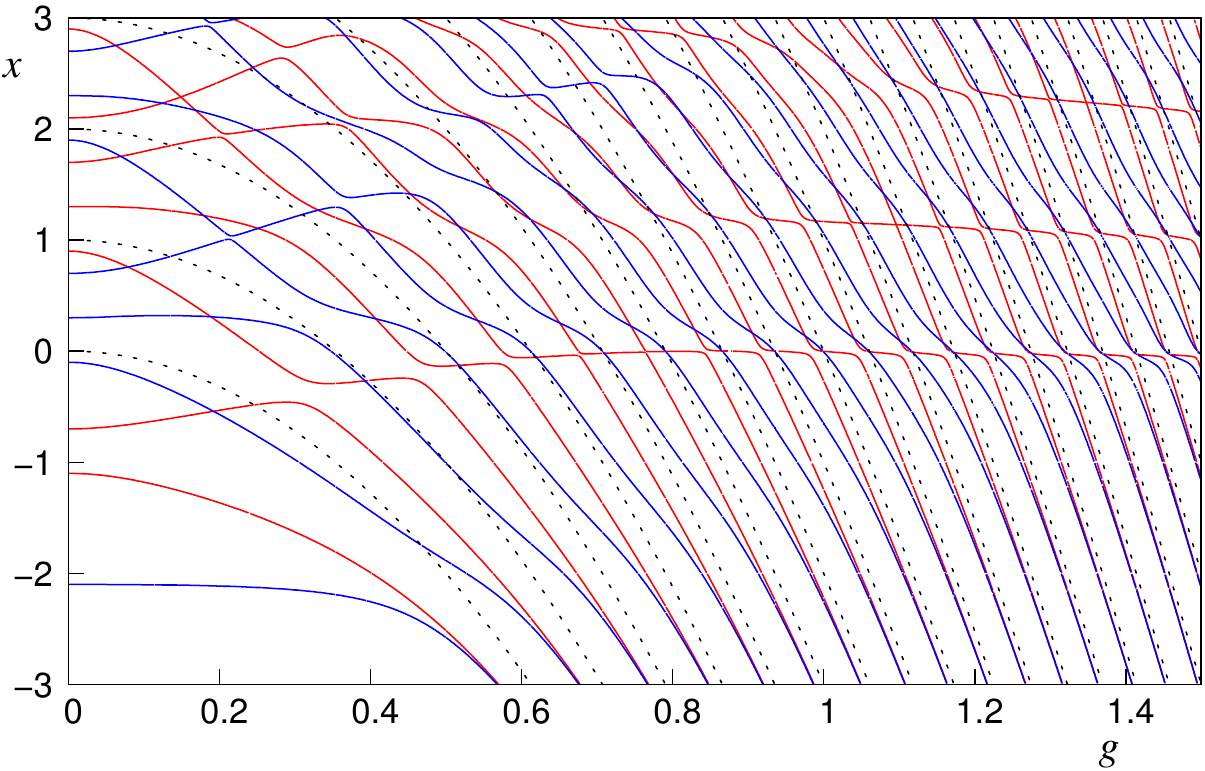}
   \caption{The spectrum of the Dicke model for even (red) and odd (blue) parity at $\Del=0.7$ and 
for varying $g$. The $y$-axis shows $x=E+g^2$.
The ground state has odd parity as in the QRM. The two ladders of eigenvalues
with different parity intersect within the regular spectrum. There are no degeneracies (but narrow avoided crossings) for fixed parity in this 
parameter window. Dashed lines denote the set $E+9g^2\in\Zpl_0$ and
emerge as limiting values in the deep strong coupling regime $g\gg 1$.
   }
    \label{dspec}
  \end{figure}

The Hilbert space of the symmetric model ADM$_3$ with $g_i'\equiv g'=\sqrt{3}g$, $\om_{0i}\equiv \om_0=2\Del$ may be splitted according to 
$\frac{1}{2}\otimes\frac{1}{2}\otimes\frac{1}{2}=
\frac{1}{2}\oplus\frac{1}{2}\oplus\frac{3}{2}$. Each of the two spin-1/2 components are equivalent to the QRM, while the last component describes a single spin-3/2 coupled to the radiation mode with Hamiltonian,
 \beq
H_{D3}=\ad a + 2\Del\Jz + 2g(a +\ad)\Jx,
\label{hamd3}
\eeq
where $\Jz$ and $\Jx$ are generators of $SU(2)$ in the (four-dimensional) spin-$3/2$ representation. The $\Zz_2$-generator has here the form $\Pr=\exp(i\pi\ad a)\R$. $\R$ is an involution acting in spin-space as  $\R\Jz\R=\Jz$, $\R\Jx\R=-\Jx$. 
Application of $\Pr$ gives the following differential operator in each parity subspace,
\beq
H_\pm=z \dd 
+\Del
\left(
\begin{array}{cc}
0&\sqrt{3}\\
\sqrt{3}&\pm2\T
\end{array}
\right)
-g
\left(
\begin{array}{cc}
3&0\\
0&1
\end{array}
\right)\left(\dd + z\right).
\label{hapm}
\eeq
Employing now the same machinery as in the previous section, we obtain again four coupled first order equations having regular singular points at $\pm g$, $\pm 3g$ and an irregular singular point (s-rank two) at infinity \cite{b-dicke}. 
Because of the stronger symmetry of \eqref{hamd3} compared to \eqref{hamd2}, the matrix $M(E)_\pm$, whose determinant gives the $G$-function is only $6\times6$.
It contains 24 functions evaluated at the points $z_0=2g$ and $z_0'=0$. The poles of $G_\pm(E)$ (giving $\s^{nd}_{exc}$) are located at $E+g^2\in\Zpl$ and $E+9g^2\in\Zpl_0$. The curves determined by the latter set are also the limiting values for the spectrum for very large coupling.
Fig.~\ref{dspec} shows the spectral graph of the model as function of the coupling $g$. As in the $N=2$ case, the degeneracies occur within the regular spectrum between states of different parity. However, the fact that we have now a determinant as $G$-function means that in principle degeneracies within the {\it same} parity chain are not excluded as the corresponding matrix $M_\pm(E)$ could have a higher-dimensional kernel at a specific energy.
Up to now there is no numerical evidence for this scenario and the existence of these novel degeneracies is an open question.
%
%
%
%
%
\section{Conclusions}\label{sec-concl}
We have seen in the previous sections that the classical theory of linear differential equations in the complex domain may be used to solve exactly  elementary but important problems in the field of theoretical and applied physics, contributing in this way to a better understanding of the basic models of circuit QED, which has been envisioned as a promising environment to implement devices capable of performing quantum computations.  

The mathematical technique relies on the Bargmann Hilbert space $\B$, which allows  to represent the Hamiltonians from Quantum Optics as differential operators acting on functions of a complex variable $z$. Of central importance is here Bargmann's two-fold spectral condition, which demands not only normalizability  with respect to the scalar product but also analyticity in $\Cc$ for any function $\p(z)$ being an element of $\B$. In this way it becomes possible to use the easily accessible singularity structure of the corresponding differential equations to implement the spectral condition {\it without} recourse to a polynomial ansatz for the wave functions, which works for elementary integrable systems like the harmonic oscillator, the hydrogen atom and the Jaynes-Cummings model, but fails already for the quantum Rabi model, which is nevertheless integrable in a well-defined sense \cite{b-prl} (for a recent comparison with Yang-Baxter integrability see \cite{bat}).

The method can be extended to models of central relevance for quantum technology, the Dicke models with a small number of qubits. These models are no longer integrable in view of the level labeling criterion \cite{b-prl} but exactly solvable with the presented technique, although many characteristic simplifications of the quantum Rabi model are absent. Further applications concern models with a single irregular singular point as the two-photon Rabi model \cite{trav}, or the anharmonic oscillator. One may also try to extend the formalism to multi-mode models \cite{fan}. Notwithstanding these generalizations, we note that the already solved systems give rise to a set of open mathematical problems like the conjecture on the level distribution presented in section \ref{subsec-reg}, or the question whether a novel class of degeneracies exists in the Dicke models (section \ref{sec-d3}). Thus, a future research direction will be the exploration of the recently observed connection \cite{wak2} between the quantum Rabi model and the non-commutative harmonic oscillator \cite{ichi}. It should be possible to transfer mathematical methods used for the study of the latter to the problems mentioned above. The techniques used in \cite{hiro,wak3, hiros,hiro2} could be applied f.e. to investigate the level crossing appearing in the ground state of the 
anisotropic quantum Rabi model \cite{xie}.

\begin{acknowledgement}
This work was supported by Deutsche Forschungsgemeinschaft through TRR~80.
\end{acknowledgement}


\begin{thebibliography}{99}





\bibitem{cohen-tann} Cohen-Tannoudji, C., Dupont-Roc, J., Grynberg, G.:
                     Atom-Photon Interactions: 
                      Basic Processes and Applications. Wiley, Weinheim (2004)

\bibitem{haroche}   Haroche, S., Raymond, J.M.:
                    Exploring the Quantum. Oxford Univ. Press, New York (2006)  
                   
\bibitem{r-gates}   Romero, G., Ballester, D., Wang, Y.M., 
                    Scarani, V., Solano, E.:
                   Ultrafast Quantum Gates in Circuit QED.
                    Phys. Rev. Lett. {\bf 108}, 120501 (2012)

\bibitem{nielsen}   Nielsen, M.A., Chuang, I.L.:
                    Quantum Computation and Quantum Information. 
                    Cambridge Univ. Press, Cambridge (2000)    


\bibitem{wallr} Wallraff, A., Schuster, D.I., Blais, A., Frunzio, L., Huang, 
                R.S., Majer, J.,
                Kumar, S., Girvin, S.M., Schoelkopf, R.J.:
               Strong coupling of a single photon 
               to a superconducting qubit using 
               circuit quantum electrodynamics.
                Nature {\bf 431}, 162 (2004)

\bibitem{niemc} Niemczyk, T., Deppe, F., Huebl, H., Menzel, E.P., Hocke, F.:
               Circuit quantum electrodynamics in the 
               ultrastrong-coupling regime. 
                Nature Physics {\bf 6}, 772 (2010)

\bibitem{forn} Forn-Diaz, P., Lisenfeld, J., Marcos D., 
               Garc\'ia-Ripoll, J.J., Solano, E., 
               Harmans, C.J.P.M., Mooij, J.E.:
              Observation of the Bloch-Siegert shift in a 
              qubit-oscillator system in the
              ultrastrong coupling regime.   
               Phys. Rev. Lett. {\bf 105}, 237001 (2010)

\bibitem{rabi}  Rabi, I.I.:
               On the Process of Space Quantization.
                Phys. Rev. {\bf 49}, 324 (1936)


\bibitem{JC}  Jaynes, E.T., Cummings, F.W.: 
             Comparison of Quantum and Semiclassical Radiation Theories
             with Application to the Beam Maser.  
              Proc. IEEE {\bf 51}, 89 (1963)

\bibitem{klimov} Klimov, A.B., Sainz, I., Chumakov, S.M.:
                Resonance expansion versus rotating-wave approximation. 
                 Phys. Rev. A {\bf 68}, 063811 (2003)

\bibitem{osterloh} Amico, L., Frahm, H., Osterloh, A., Ribeiro, G.A.P.: 
                  Integrable spin-boson models descending from 
                  rational six-vertex models.
                   Nucl. Phys. B {\bf 787}, 283 (2007)

\bibitem{arnold}  Arnold, V.I.:
                  Mathematical Methods of Classical Mechanics. 
                  Springer-Verlag, New York (1989)
 
\bibitem{caux}    Caux, J.S., Mossel, J.:
                 Remarks on the notion of quantum integrability.    
                  J. Stat. Mech. P02023 (2011)

\bibitem{sol1} Casanova, J., Romero, G., Lizuain, I., Garc\'ia-Ripoll, 
               J.J., Solano, E.: 
              Deep Strong Coupling Regime of the Jaynes-Cummings Model. 
               Phys. Rev. Lett. {\bf 105}, 263603 (2010)
 
\bibitem{b-prl}  Braak, D.:
                Integrability of the Rabi Model.
                 Phys. Rev. Lett. {\bf 107}, 100401 (2011) 

\bibitem{barg} Bargmann, V.:
              On a Hilbert space of analytic functions and
              an associated integral transform, part I.  
               Commun. Pure Appl. Math. {\bf 14}, 187 (1961)

\bibitem{b-contfrac} Braak, D.: 
              Continued fractions and the Rabi model.     
               J. Phys. A: Math.Theor. {\bf 46}, 175301 (2013)

\bibitem{ince} Ince, E.L.:
               Ordinary Differential Equations. Dover, New York (1956)

\bibitem{slav} Slavyanov, S.Y., Lay, W.:
               Special Functions. A Unified Theory Based on Singularities.
               Oxford Univ. Press, New York (2000) 

\bibitem{b-wronski} Braak, D.: 
                   A generalized $G$-function for the quantum Rabi model.
                    Ann. Phys. (Berlin) {\bf 525}, L23 (2013)

\bibitem{zhong}  Zhong, H., Xie, Q., Batchelor, M.T., Lee, C.:
                 Analytical eigenstates for the quantum Rabi model.
                 J. Phys. A: Math. Theor. {\bf 46}, 14 (2013)  


\bibitem{ronv}      Ronveaux, A. (Ed.):
                    Heun's Differential Equations.
                    Oxford Univ. Press, New York (1995) 

\bibitem{mac}       Maciejewski A.J., Przybylska, M., Stachowiak, T.:
                    Full spectrum of the Rabi model.
                    Phys. Lett. A {\bf 378}, 16 (2014)

\bibitem{turb}      Turbiner, A.V.:
                    Quasi-Exactly-Solvable Problems and $sl(2)$ Algebra.      
                    Commun. Math. Phys. {\bf 118}, 467 (1988)

\bibitem{wak1}      Wakayama, M., Yamasaki, T.:
                   The quantum Rabi model and the Lie algebra representations 
                    of $\mathfrak{sl}_2$. 
                    J. Phys. A: Math. Theor. {\bf 47}, 335203 (2014)


\bibitem{schweb} Schweber, S.:
                On the Application of Bargmann Hilbert Spaces
                to Dynamical Problems.   
                 Ann. Phys., NY {\bf 41}, 205 (1967)

\bibitem{gautschi} Gautschi, W.:
               Computational aspects of three-term recurrence relations.
                   SIAM Review {\bf 9}, 24 (1967)

\bibitem{swain} Swain, S.:
               A continued fraction solution to the problem of a 
               single atom interacting with a single radiation mode 
               in the electric dipole approximation.
                J. Phys. A: Math. Nucl. Gen. {\bf 6}, 192 (1973)

\bibitem{tur}   Tur, E.A.:
               Energy Spectrum of the Hamiltonian of the Jaynes-Cummings 
               Model without Rotating-Wave Approximation. 
                Opt. Spectrosc. {\bf 91}, 899 (2001)              


\bibitem{dicke} Dicke, R.H.:
               Coherence in Spontaneous Radiation Processes.
                Phys. Rev. {\bf 93}, 99 (1954)

\bibitem{tavis} Tavis, M., Cummings, F.W.:
               Exact Solution for an N-Molecule-Radiation-Field Hamiltonian.
                Phys. Rev. {\bf 170}, 379 (1968)

\bibitem{haack}  Haack, G,, Helmer, F., Mariantoni, M., Marquardt, F., 
                 Solano, E.:
                Resonant quantum gates in circuit quantum electrodynamics.
                 Phys. Rev. B {\bf 82}, 024514 (2010)

\bibitem{peng}   Peng, J., Ren, Z.Z., Braak, D., Guo, G.J.,  
                 Ju, G.X.,  Zhang, X., Guo, X.Y.:
                Solution of the two-qubit quantum Rabi model and its  
                exceptional eigenstates.
                 J. Phys. A: Math. Theor. {\bf 47}, 265303 (2014) 


\bibitem{rod}   Chilingaryan, S.A., Rodr\'iguez-Lara, B.M.:
                The quantum Rabi model for two qubits.
                J. Phys. A: Math. Theor. {\bf 46}, 335301 (2013) 

\bibitem{b-dicke} Braak, D.: 
                  Solution of the Dicke model for $N=3$. 
                  J. Phys. B: At. Mol. Opt. Phys. {\bf 46}, 224007 (2013)

\bibitem{bat}       Batchelor, M.T., Zhou, H.Q.:
                    Integrability versus exact solvability   
                    in the quantum Rabi and Dicke models.
                    Phys. Rev. A, \textbf{91}, 053808 (2015)

\bibitem{trav}     Trav{\v e}nec, I.:
                   Solvability of the two-photon Rabi Hamiltonian.
                   Phys. Rev. A, {\bf 85}, 043805 (2012)

                
\bibitem{fan}      Fan, J., Yang, Z., Zhang, Y., Ma, J., Chen, G., Jia, S.:
                   Hidden continuous symmetry and Nambu-Goldstone mode in 
                   a two-mode Dicke model.
                   Phys. Rev. A, {\bf 89}, 023812 (2014) 

\bibitem{wak2}     Wakayama, M.:
                   Equivalence between the eigenvalue problem of non-commutative
                   harmonic oscillators and existence of holomorphic solutions
                   of Heun differential equations, eigenstates degeneration
                   and the Rabi model.
                   MI-preprint series, Kyushu University (2013)


\bibitem{ichi}    Ichinose, T., Wakayama, M.:
                  Zeta Functions for the Spectrum of the Non-Commutative
                  Harmonic Oscillators. 
                  Commun. Math. Phys. {\bf 258}, 697 (2005) 

\bibitem{hiro}    Hirokawa, M.:
                  The Dicke-type crossing among eigenvalues of 
                  differential operators in a class of non-commutative 
                  oscillators.
                  Indiana Univ. Math. J. {\bf 58}, 1493 (2009)

\bibitem{wak3}    Wakayama, M.:
                  Simplicity of the lowest eigenvalue of non-commutative
                  harmonic oscillators and the Riemann scheme of a certain
                  Heun's differential equation.
                  Proc. Japan Acad. {\bf 89} Ser. A, 69 (2013)
   
\bibitem{hiros}   Hiroshima, F., Sasaki, I.:
                  Spectral analysis of non-commutative harmonic oscillators:
                  The lowest eigenvalue and no crossing.
                  J. Math. Anal. Appl. {\bf 105}, 595 (2014)

 \bibitem{hiro2}   Hirokawa, M., Hiroshima, F.:
                  Absence of energy level crossing for the ground state energy
                  of the Rabi model.
                  Comm. Stoch. Anal. (to appear)

\bibitem{xie}     Xie, Q.T., Cui, S., Cao, J.P., Amico, A., Fan, H.:
                  Anisotropic Rabi model.
                  Phys. Rev. X {\bf 4}, 021046 (2014)




\end{thebibliography}
\end{document}